
\input harvmac
\input epsf

\def\latilde{\tilde\lambda}
\def\rhotilde{\tilde\rho}

\def\Atilde{\tilde A}

\def\threebar{{\overline{3}}}
\def\5bar{{\overline{5}}}

\def\sy{supersymmetry}

\def \in{\leftskip = 40 pt\rightskip = 40pt}
\def \out{\leftskip = 0 pt\rightskip = 0pt}

\def\npb{{Nucl.\ Phys.\ }{\bf B}}

\def\plb{{Phys.\ Lett.\ }{\bf B}}
\def\prd{{Phys.\ Rev.\ }{\bf D}}
\def\prl{Phys.\ Rev.\ Lett.\ }

\def\lf{16\pi^2}

{\nopagenumbers
\line{\hfil LTH 352}
\line{\hfil hep-ph/9506467}
\vskip .5in
\centerline{\titlefont Infra--red soft universality}
\vskip 1in
\centerline{\bf P.M.~Ferreira, I.~Jack, and D.R.T.~Jones}
\bigskip
\centerline{\it DAMTP, University of Liverpool, Liverpool L69 3BX, U.K.}
\vskip .3in
We show that in a special class of theories
the commonly assumed universal form of
the soft \sy --breaking terms
is  approached in the infra--red limit. The resulting universal scalar
mass and trilinear coupling are predicted in terms of the gaugino mass.
\Date{June 1995}}

The  supersymmetric standard model (SSM) has gained widespread
acceptance as a framework for physics at and above 1~TeV, encouraged by
the unification of gauge couplings within  this theory at a scale of
$M_U=10^{16}$~GeV. It is desirable that the  soft breaking couplings
should adopt a roughly ``universal'' form  at $M_U$;
significant deviations from universality would give rise to
flavour-changing neutral currents in the theory at the weak scale. It
happens that popular scenarios which explain the soft-breaking terms  as
generated by supersymmetry breaking in the ``hidden sector'' of an
underlying supergravity theory (possibly ultimately arising from string
theory) do in fact make at least plausible  a universal form of the kind
required. However, this universal form would pertain at or near the
Planck scale $(M_P)$  and in general diversions away from universality
would be expected as the couplings evolved down to $M_U$
\ref\Pol{N.~Polonsky and A.~Pomarol, \prl 73 (1994) 2292.}. Although
this  need not be disastrous for phenomenology,  it does mean that low
energy predictions are sensitive to the  nature of the unified theory;
and the problem of a fully convincing  explanation of the origin of
universality at $M_P$ remains unsolved.

In a recent paper \ref\jjb{I.~Jack and D.R.T.~Jones, \plb 349 (1995)
294.} we showed that if the dimensionless  couplings obeyed a certain
relation (which we shall call generically the $P={1\over3}Q$ condition),
then a particular universal form for the soft-breaking couplings was
preserved by the renormalisation group evolution down to $M_U$.
Moreover, this universal form is in fact predicted  by a fairly generic
superstring scenario in which supersymmetry breaking is  engendered by
dilaton and ``size modulus'' vevs. This is all well and good,  but
suffers from the drawback that there is no obvious reason for string
theory to yield dimensionless couplings satisfying the required
$P={1\over3}Q$ constraint. A much more interesting hypothesis, it seems
to us, is the following:  if the grand unified theory above $M_U$ is
such that dimensionless  couplings can in principle be found to satisfy
the $P={1\over3}Q$ condition, then such a configuration of dimensionless
couplings may represent  an attractive  infra-red (IR) fixed point,
which can be approached quite closely as the theory evolves towards
$M_U$. Moreover the RG-invariant universal form for the soft-breaking
couplings  alluded to above may also then constitute an IR fixed point
which again may be approached quite closely at $M_U$.  This would mean
that we would no longer have to impose universality  at $M_U$, or $M_P$;
for a wide range of possible input parameters the  unified theory would
evolve towards universality at $M_U$.   We have pursued the
phenomenological consequences of this idea elsewhere  \ref\jjra{I.~Jack,
D.R.T.~Jones and K.L.~Roberts, LTH 347
(hep-ph/9505242)}\ref\jbalt{D.R.T.~Jones, LTH 350 (hep-ph/9506206).}; in
this  paper we explore the conditions in which it may be realised at
$M_U$  using a number of toy models.

One might be tempted to argue that there
is not a sufficient range of energy between the Planck scale, $M_P$,
and $M_U$ for
any significant progress towards a fixed point to take place. However,
as was recently pointed out by Lanzagorta and Ross\ref\lanz{
M.~Lanzagorta and G.G.~Ross, \plb 349 (1995) 319.}, the rate of evolution
of couplings in the unified theory is enhanced relative to the SSM
by the larger field content and by the potential lack of asymptotic freedom.
Of course how precisely universality is approached will depend on the model.

We start by reviewing our previous work and then demonstrate the attractive
nature of our fixed points in a restricted case. We then discuss
in detail a rather more realistic $(SU_3 )^3$ grand unified model which admits
the $P={1\over3}Q$ condition and present some numerical results which
display the approach to universality at $M_U$.
We begin with our results for a general theory.
The Lagrangian $L_{\rm SUSY} (W)$ is defined by the superpotential
\eqn\Ea{
W={1\over6}Y^{ijk}\Phi_i\Phi_j\Phi_k+{1\over2}\mu^{ij}\Phi_i\Phi_j. }
$L_{\rm SUSY}$ is the Lagrangian for  the $N=1$ supersymmetric
gauge theory, containing the gauge multiplet ($\lambda$ being the
gaugino) and a chiral superfield $\Phi_i$ with component fields
$\{\phi_i,\psi_i\}$ transforming as a
representation $R$ of the gauge group $\cal G$.
We assume that there are no gauge-singlet
fields and that $\cal G$ is simple. (The generalisation to a semi-simple
group is trivial.)
The soft breaking is incorporated in $L_{\rm SB}$, given by
\eqn\Eb{
L_{\rm SB}=(m^2)^j_i\phi^{i}\phi_j+
\left({1\over6}h^{ijk}\phi_i\phi_j\phi_k+{1\over2}b^{ij}\phi_i\phi_j
+ {1\over2}M\lambda\lambda+{\rm h.c.}\right)}
(Here and elsewhere, quantities with superscripts are complex conjugates of
those with subscripts; thus $\phi^i\equiv(\phi_i)^*$.)
The superpotential $W$ undergoes no infinite renormalisation
so that we have, for instance
\eqn\Ec{
\beta_Y^{ijk}=Y^{ijp}\gamma^k{}_p+(k\leftrightarrow
i)+(k\leftrightarrow j),}
where $\gamma$ is the anomalous dimension for $\Phi$.
The one-loop results for the gauge coupling $\beta$-function $\beta_g$ and
for $\gamma$ are given by
\eqn\Ed{
\lf\beta_g^{(1)}=g^3Q,\quad\hbox{and}\quad
\lf\gamma^{(1)i}{}_j=P^i{}_j,}
where
\eqna\Ee$$\eqalignno{
Q&=T(R)-3C(G),\quad\hbox{and}\quad &\Ee a\cr
P^i{}_j&={1\over2}Y^{ikl}Y_{jkl}-2g^2C(R)^i{}_j. &\Ee b\cr}$$
Here
\eqn\Ef{
T(R)\delta_{AB} = \Tr(R_A R_B),\quad C(G)\delta_{AB} = f_{ACD}f_{BCD}
\quad\hbox{and}\quad C(R)^i{}_j = (R_A R_A)^i{}_j.}
The one-loop $\beta$-functions
for the soft-breaking couplings are given by
\eqna\Eg$$\eqalignno{
\lf\beta_h^{(1)ijk}&=h^{ijl}P^k{}_l+Y^{ijl}X^k{}_l \quad
+ (k\leftrightarrow i) + (k\leftrightarrow j) &\Eg a\cr
\lf[\beta_{m^2}^{(1)}]^j{}_i&=
{1\over2}Y_{ipq}Y^{pqn}(m^2)^j{}_n+{1\over2}Y^{jpq}Y_{pqn}(m^2)^n{}_i
+2Y_{ipq}Y^{jpr}(m^2)^q{}_r\cr &\quad
+h_{ipq}h^{jpq}-8g^2MM^* C(R)^j{}_i,&\Eg b\cr
\lf\beta_b^{(1)ij}&=b^{il}P^j{}_l
+\mu^{il}X^j{}_l \quad + (i\leftrightarrow j),&\Eg c\cr
\lf\beta_M^{(1)}&=2g^2QM,&\Eg d\cr}$$
where
\eqn\Ei{
X^i{}_j=h^{ikl}Y_{jkl}+4Mg^2C(R)^i{}_j}
and we have dropped a $\tr[R_Am^2 ]$ term in Eq.~\Eg{b}\
because ${\cal G}$ is simple,
and terms of the type $Y_{ijk}b^{jk}$ and $Y_{ijk}\mu^{jk}$
because there are no gauge singlets.
We then showed in Ref.~\jjb\ that the conditions
\eqna\Ek
$$
\eqalignno{h^{ijk}&=-MY^{ijk},&\Ek a\cr
(m^2)^i{}_j&={1\over3}MM^*\delta^i{}_j,&\Ek b\cr
b^{ij}&=-{2\over3}M\mu^{ij}&\Ek c\cr}
$$
are RG invariant at one loop provided we
impose the $P={1\over3}Q$ condition
\eqn\El{
P^i{}_j = g^2P\delta^i{}_j = {1\over3}g^2Q\delta^i{}_j.}
Moreover, the condition Eq.~\El\ is itself RG invariant up to
at least two loops.
In other words, dimensionless couplings satisfying Eq.~\El\ and soft-breaking
couplings satisfying Eq.~\Ek{}\ represent fixed points of the RG evolution; it
remains to confirm our claim that they can be
IR attractive.\footnote{\dag}{In the special case of a finite theory,
we have $P = Q = 0$, and soft breakings satisfying Eq.~\Ek{}\ preserve
finiteness
\ref\jmy{D.R.T.~Jones, L.~Mezincescu and Y.-P.~Yao, \plb148 (1984) 317.}
\ref\jj{I.~Jack and D.R.T.~Jones, \plb333 (1994) 372.}.
For the $N=4$ case, the fact that these results for the soft terms are
approached in the IR limit was pointed out in Ref.~\ref\marty{M.B.~Einhorn,
G.~Goldberg and E.~Rabinovici, \npb 256 (1985) 499.}.}
We shall do this analytically in a somewhat restricted case but our
numerical experience with a more complex example indicates that this property,
while not completely general, is at least a plausible feature of a realistic
theory.
Consider then the case of a theory with fields $\phi^i$ in an irreducible
representation of $\cal G$, for which
\eqn\Em{C(R)^i{}_j=C(R)\delta^i{}_j, \qquad Y^{ikl}Y_{jkl}=Y\delta^i{}_j }
so that $P^i{}_j=g^2 P\delta^i{}_j$, where $P = Y/(2g^2) - 2C(R)$.
It is easy to show that the standard
fixed point\ref\Pendleton{B.~Pendleton and G.G.~Ross,
\plb 98 (1981) 291.}
in the evolution of the ratio of the Yukawa
to gauge couplings corresponds to $P= {1\over 3}Q$,
and that it exists as long as $Q + 6C(R) > 0$.

 Suppose further that we
have soft-breaking couplings given by
\eqna\En
$$
\eqalignno{h^{ijk}&=-x MY^{ijk},&\En a\cr
(m^2)^i{}_j&= yMM^*\delta^i{}_j.&\En b\cr}
$$
It is easy to show using Eqs.~\Eg{}-\Ei\ and Eq.\Em\ that at
the fixed point $P={1\over3}Q$ we have
\eqn\Eo{
\lf\beta_{x}=12(x-1)C(R)g^2,}
so that $x=1$ is an IR fixed point.
Then with $P={1\over3}Q$ and $x=1$,
\eqn\Ep{
\lf\beta_{y}=2(y- {1\over 3})[6C(R)-Q]g^2 }
so that $y={1\over 3}$ is also an IR fixed point, as long as
$6C(R) - Q > 0$. Finally if we suppose that the representation $R$
also permits a quadratic invariant and set
\eqn\Ep{
b^{ij} = -{2\over 3}z M\mu^{ij},}
we find that $z = 1$ is IR-attractive as long as $Q<0$.

In more complicated cases, it can happen that while there does  exist an
IR--attractive fixed point for the dimensionless couplings,  it does not
correspond to $P = {1\over 3}Q$. The trilinear scalars will then still
have the fixed point  corresponding to  Eq.~\Ek{a}.  This fixed point
may or may not be  IR attractive, however.  In these cases neither
Eq.~\Ek{b}\ nor Eq.~\Ek{c}\ will correspond in general to fixed points.
Although the scalar mass evolution may still exhibit fixed point
behaviour, this will not correspond to a {\it common\/} mass,  as we
have in Eq.~\Ek{b}. It is for this reason that we favour theories which
can satisfy $P = {1\over 3}Q$. It may also be that given a theory
admitting $P = {1\over 3}Q$,  the behaviour of the Yukawa couplings may
be governed (for large  initial values at $M_P$) by  quasi--fixed
point\lanz\ref\hill{C.T.~Hill, \prd 24 (1981) 691\semi C.T.~Hill,
C.N.~Leung and S.~Rao, \npb 262 (1985) 517.}  rather than fixed--point
behaviour. We should emphasise that in order  to realise our goal of
soft universality the Yukawa couplings must approach  the actual $P =
{1\over 3}Q$ fixed point.
It can also happen that while $P = {1\over 3}Q$ is indeed IR attractive,
one or more of the conditions in Eq.~\Ek{}\ are saddle points. In specific
models where this is the case, it can happen that for quite reasonable
regions of parameter space the RG trajectories approach quite close
to the saddle point when integrated from $M_P$ to $M_U$. We will see an
example of  this later.

As our semi--realistic model we will take
$SU_3\otimes SU_3\otimes SU_3$.  $(SU_3)^3$
is a maximal subgroup of $E_6$; both
groups  have attractive features as candidate GUTs, particularly in the
string context. Here we consider the basic  case of  a $(SU_3)^3$
theory with $n$ sets each of the representations  $X\equiv (3, 3, 1)$,
$Y\equiv (1, \threebar ,  3)$ and $Z\equiv (\threebar , 1, \threebar )$.
The superpotential for the theory is :
\eqn\Eq{
W = {1\over{3!}}(\lambda_1 X^3 + \lambda_2 Y^3  + \lambda_3 Z^3 )
+ \rho XYZ.}
Here $\lambda_1 X^3\equiv ( \lambda_1 )^{\alpha\beta\gamma}
X_{\alpha}X_{\beta}X_{\gamma}$, where $\alpha, \beta\cdots = 1\cdots n$.
If we set the three gauge couplings all equal to $g$ then it is easy to see
that they remain equal under renormalisation, and
$Q=3n-9$. (We may choose to imagine other sectors of the theory
also contributing to $Q$, in which case $Q$ becomes a free parameter,
subject only to $Q > 3n - 9$.) The $P={1\over 3}Q$ condition
for this theory consists of the set of  equations:
\eqn\Er{
(2\lambda_i^2 + 3\rho^2)^{\alpha}_{\beta} = {1\over 3}(16 + Q)g^2
\delta^{\alpha}_{\beta} , \quad i=1\cdots 3\quad (\hbox{no sum on $i$}).}
where $(\lambda_i^2)^{\alpha}_{\beta} =
(\lambda_i)^{\alpha\gamma\delta} (\lambda_i{}^* )_{\beta\gamma\delta .}$
It is easy to see that these conditions are identical to those
obtained by requiring the  Yukawa couplings to be at the PR fixed points.
Notice that in this case the Yukawa couplings are not
completely determined by the $P = {1\over 3}Q$ condition.

In what follows we will suppose that we have
$(\lambda_i^2)^{\alpha}_{\beta} = (\lambda_i^2)\delta^{\alpha}_{\beta}$,
and $(\rho^2)^{\alpha}_{\beta} = \rho^2\delta^{\alpha}_{\beta}$.
Assuming also that the soft $\phi^3$ terms have the form
$( h_i )^{\alpha\gamma\delta} = A_i(\lambda_i )^{\alpha\gamma\delta}$,
and $( h_{\rho})^{\alpha\gamma\delta} = A_{\rho}\rho^{\alpha\gamma\delta}$,
then the fixed point conditions for the $A$-parameters are as follows:
\eqna\Es
$$\eqalignno{
(6\latilde_i^2 - Q)\Atilde_i + 6\Atilde_{\rho}\rhotilde^2 + 16 &= 0
\quad i=1\cdots 3,&\Es a\cr
\sum_i 2\Atilde_i\latilde_i^2
+ (9\rhotilde^2 - Q)\Atilde_{\rho} + 16 &= 0,&\Es b\cr}
$$
where we have defined  $\latilde_i = \lambda_i /g$, $\rhotilde = \rho /g$ and
$\Atilde_i = A_i / M$.  Imposing the $P = {1\over 3}Q$ condition we find
the unique fixed point  $\Atilde_i = \rhotilde = -1$, corresponding, of
course, to Eq.~\Ek{a}. The stability matrix for the four $\Atilde_i$
couplings has eigenvalues  $16-9\rhotilde^2$ (twice), $16$, and $-Q$. Thus
for IR stability we require  $Q<0$. This case is not favourable, even as
a toy model, however;  for example with  $Q=-3$ the
$P={1\over 3}Q$ point is not approached  very rapidly.  Turning to the
case $Q>0$, it is interesting that the eigenvector corresponding  to the
eigenvalue $+16$ is $(\Atilde_i, \Atilde_{\rho}) = (1, 1, 1, 1)$;  this
means that if we start at $M_P$ with $\Atilde_i = \Atilde_{\rho}$,  (or
close to it) then we will be close to our fixed point  $\Atilde_i =
\Atilde_{\rho} = -1$ at $M_U$. Thus if string theory indeed  dictates a
universal A--parameter, then even though the  fixed  point corresponding
to Eq.~\Ek{a}\ is a saddle point, it will still be IR attractive.

The fixed point for the soft $\phi\phi^*$ masses corresponding to
Eq.~\Ek{b}\ has a stability matrix with eigenvalues  $32 - 2Q -
18\rhotilde^2 (= 12\lambda_i^2)$ (twice) and $32 - 2Q$,   when
$\lambda_{1\cdots 3}$, $\rho$, and the $A$--parameters are at the fixed
point. Thus we might expect  good approach to universality for
comparatively small $\rhotilde$.

We now present some numerical  results. In our analysis we run couplings
and masses  from $M_P$ to $M_U$\footnote{\dag}{It is also possible that
above an intermediate  compactification scale $M_c$ the effective theory
contains towers  of Kaluza--Klein states; these may actually improve the
rate of  approach to the fixed point\lanz .} and look for regions of
parameter space such that  the various soft parameters  approach
 their fixed point values  with a given degree of accuracy. We use $Q =
1$ and $g(M_U)= 0.72$  throughout; with this value  the  dimensionless
couplings approach the $P={1\over 3}Q$ point for a  reasonable range of
starting values. This behaviour  is illustrated in Fig.1, where
$P-{1\over 3}Q$ is plotted against energy  scale for various starting
values of the couplings at $M_P$.  For all the curves we have $\lambda_1
= \lambda_2 = \lambda_3$, and $\lambda_i (M_P) = 5\rho (M_P)$.

Turning to the soft parameters, let us consider first what happens if
we  begin with ``weak'' universality  at $M_P$ (meaning values for the
scalar masses and $A$-parameters that  are universal, but not at the
$P={1\over 3}Q$ values). We use  parameters $x$ and $y$ as defined in
Eq.~\En{}. In Fig.2 we show how the  $A$-parameters converge,  and it
can be seen that for quite substantial regions of  parameter-space the
fixed point value $x=1$ is   approached quite closely at $M_U$.

In Fig.3 we present a similar plot showing the approach of $y$ to the fixed
point; in this case also there are sizeable regions such
that $|y-{1\over 3}|$ is small at $M_U$. Here we have taken
$\lambda_i (M_P) = \rho (M_P) = 4.9$, which is close to the
limit for perturbative believability. It is interesting that if we
take larger values of $\lambda_i (M_P), \rho (M_P)$ then although
we are then starting further from $P={1\over 3}Q$, the soft couplings
approach the fixed point more quickly; but as we increase the dimensionless
couplings, perturbation theory becomes less trustworthy, of course.

\vfill\eject

\epsfysize= 3.0in
\centerline{\epsfbox{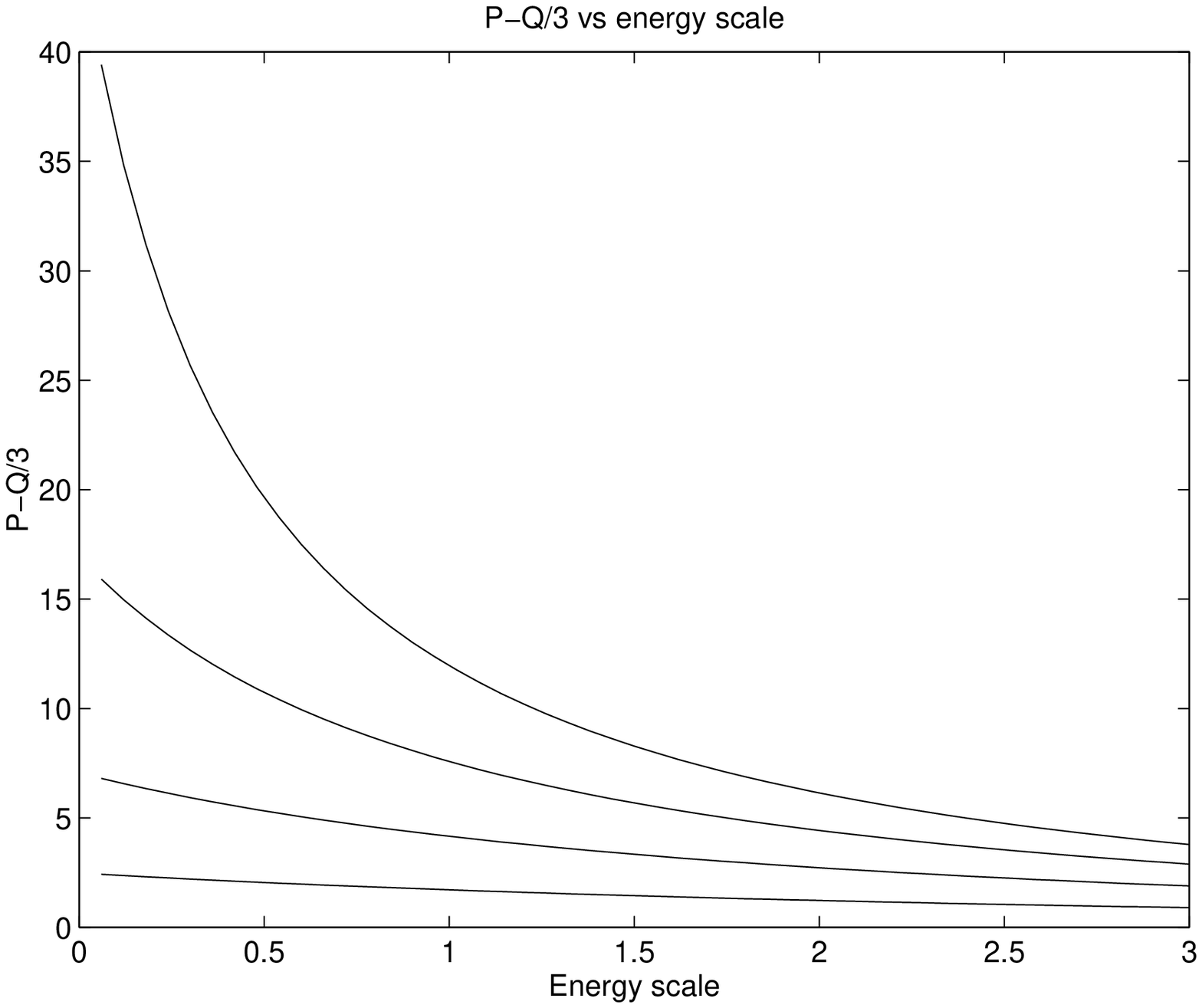}}
\in
{\it \noindent Fig.1:
Evolution of $P -{1\over 3}Q$  from $M_P$ to
$M_U$, for various input values of $\lambda_i$,  $\rho$
at $M_P$. All the curves correspond to $\lambda_i (M_P) = 5\rho (M_P)$.
The $x$-axis is
$log_{10}(M_P / {\mu})$}.
\medskip
\out

\epsfysize= 3.0in
\centerline{\epsfbox{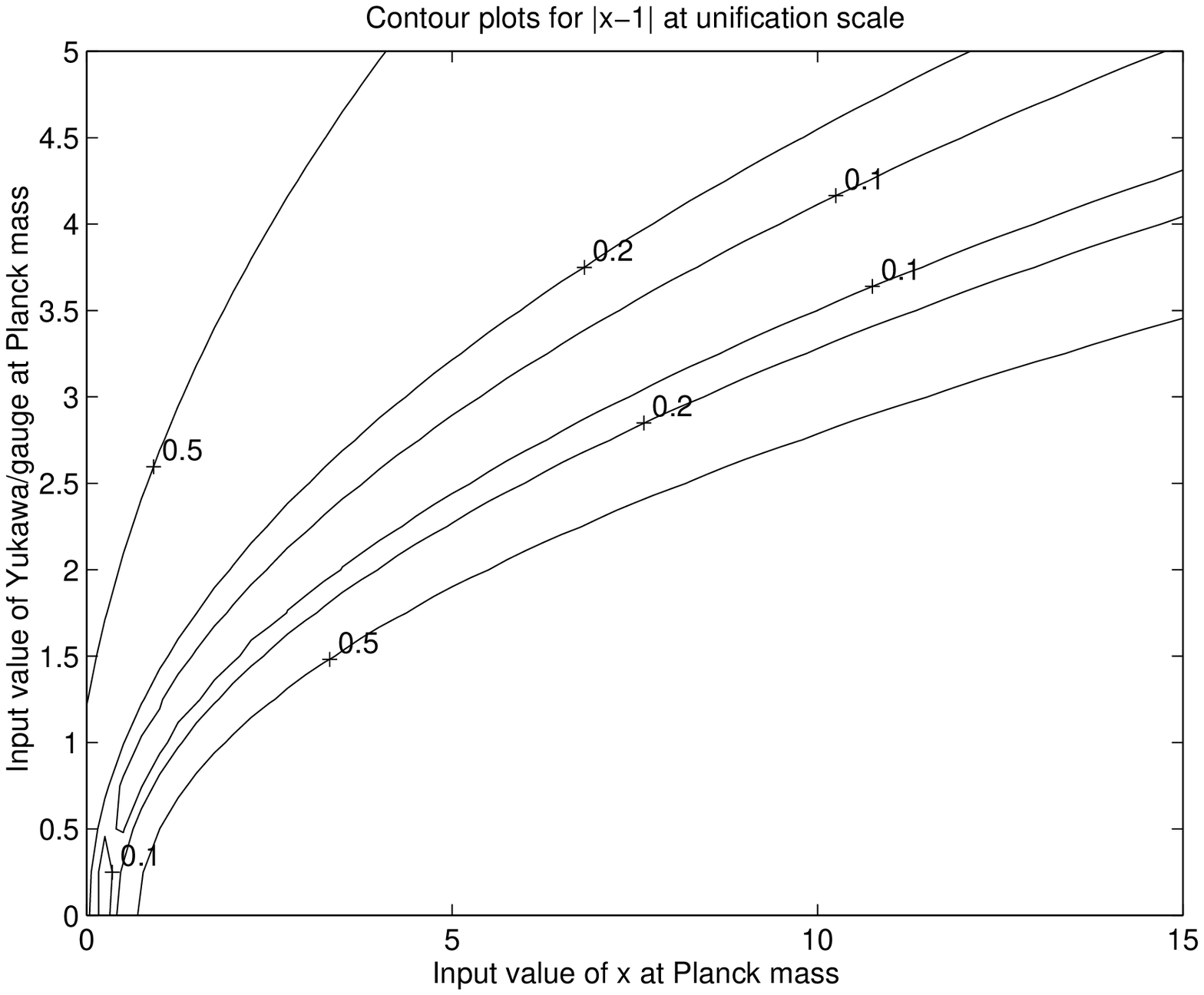}}
\in
{\it \noindent Fig.2: Contour plot showing input values of
$x$ and $\latilde_i = \rhotilde$
at $M_P$ that lead to $|x-1|< 0.1, 0.2, 0.5$ at $M_U$.}
\medskip
\out

\vfill\eject

\epsfysize= 3.0in
\centerline{\epsfbox{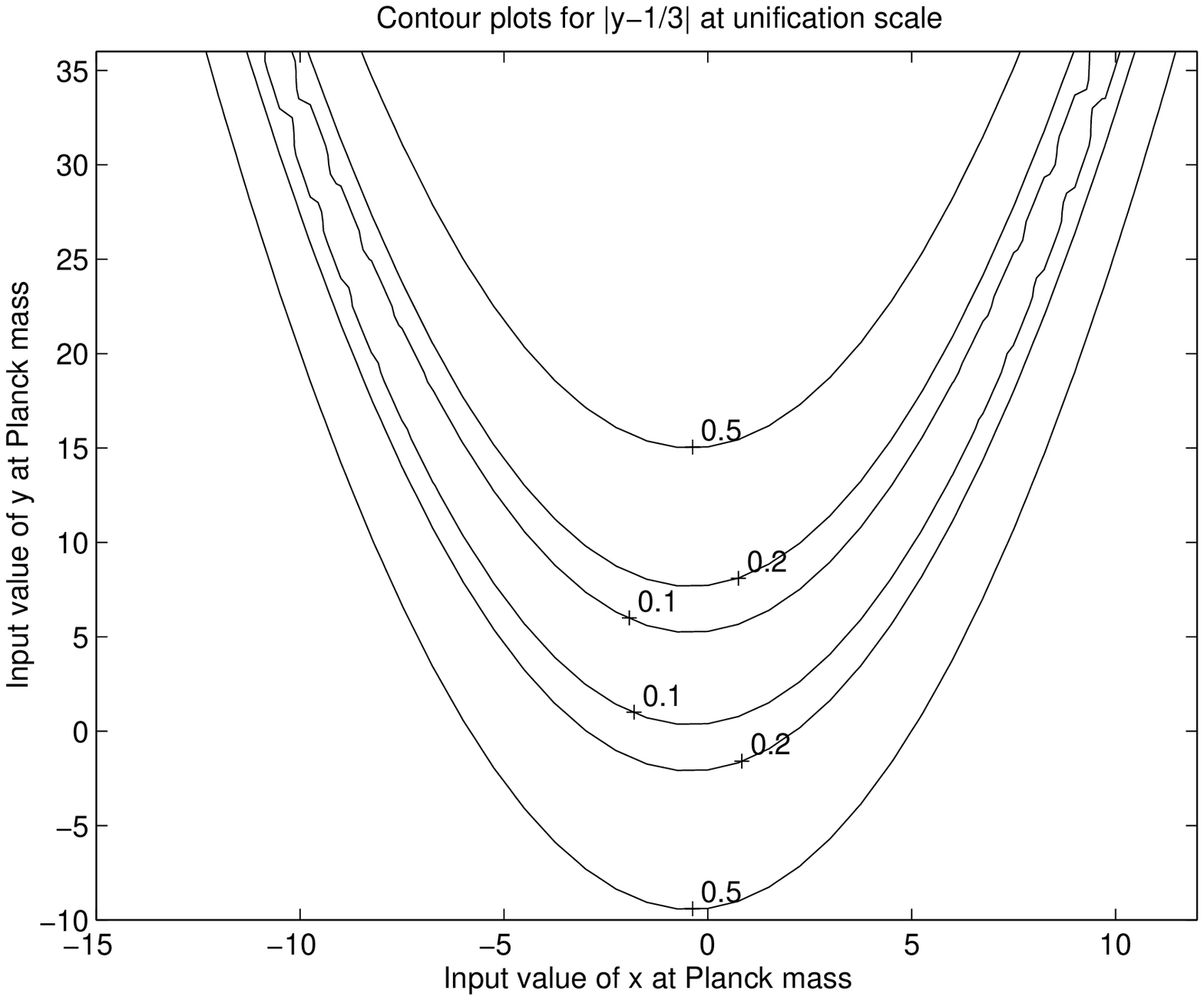}}
\in
{\it \noindent Fig.3: Contour plot showing input values of
$x$ and $y$ at $M_P$ that lead to $|y-{1\over 3}| < 0.1, 0.2, 0.5$ at $M_U$.
}
\medskip
\out

It is also interesting to explore what happens if ``weak'' universality does
not hold at the Planck mass.

\epsfysize= 3.0in
\centerline{\epsfbox{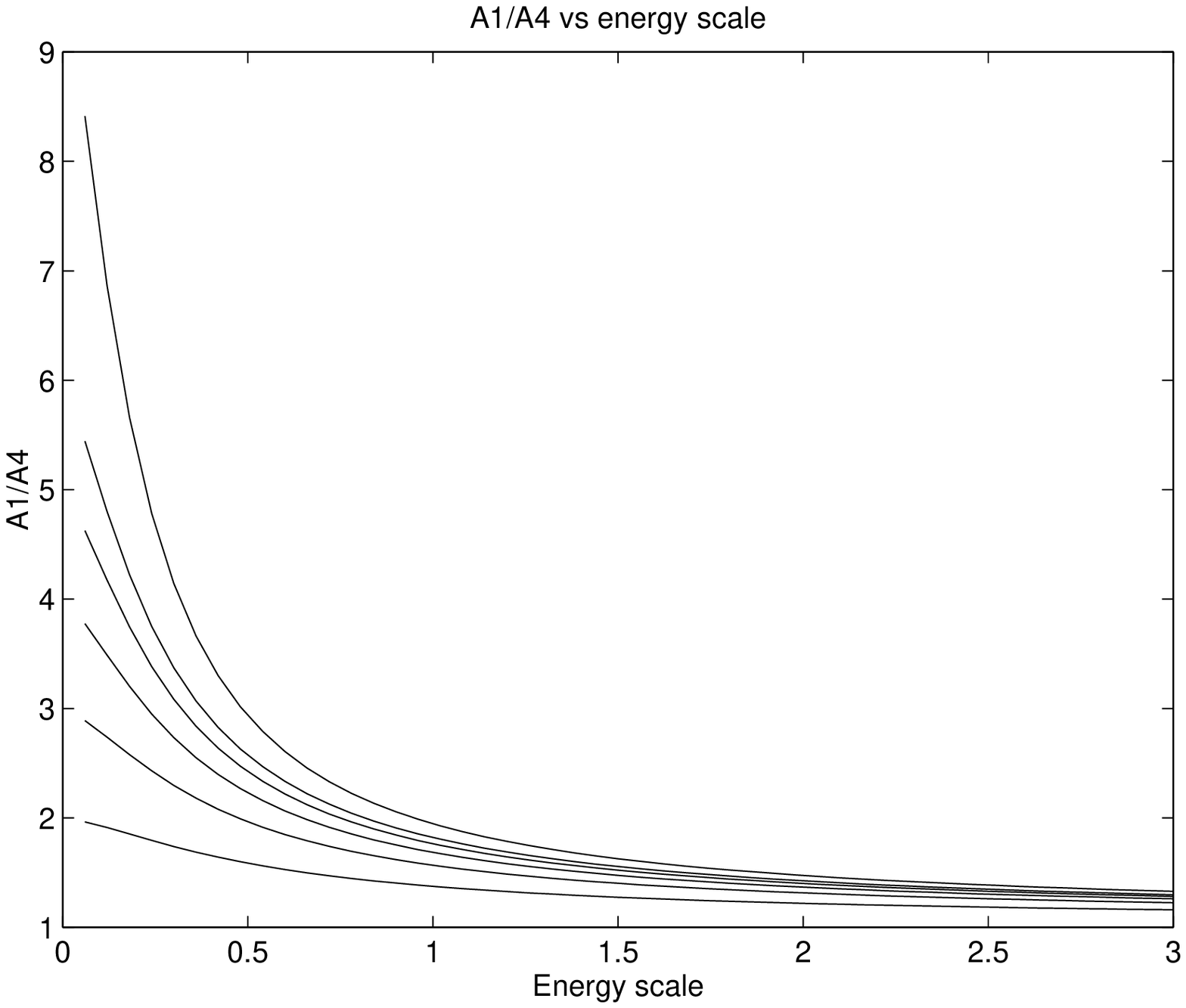}}
\in
{\it \noindent Fig.4:
Evolution of the ratio of $A_1$ to $A_{\rho} (\equiv A4)$ from $M_P$ to
$M_U$, for various input values at $M_P$. The $x$-axis is
$log_{10}(M_P / {\mu})$}.
\medskip
\out
\vfill\eject

In Fig.4 we show how the $A$-parameters evolve
if we assume that $A_1 = A_2 = A_3 = 0.1\neq A_{\rho}$ at $M_P$. Shown is the
evolution of the ratio $A_1 /A_{\rho}$ from $M_P$ to $M_U$, for various
starting values. All the curves correspond to $\lambda_i = 4.9$ and
$\rho = 0.98$. Even though, as discussed above, Eq.~\Ek{a}\ is not
IR attractive in this case, it is still approached quite well at $M_U$.

The soft scalar masses exhibit similar behaviour. In Fig.5 we show how
a non-universal choice of masses at $M_P$ converges quite
rapidly to the fixed point as we approach $M_U$.
We have taken $\Atilde_i = 0.1$, $\Atilde_{\rho}=0.02$,
$m_X = m_Y = M \neq m_Z$, and used various input values for
$(m_X / m_Z)^2$.

\epsfysize= 3.0in
\centerline{\epsfbox{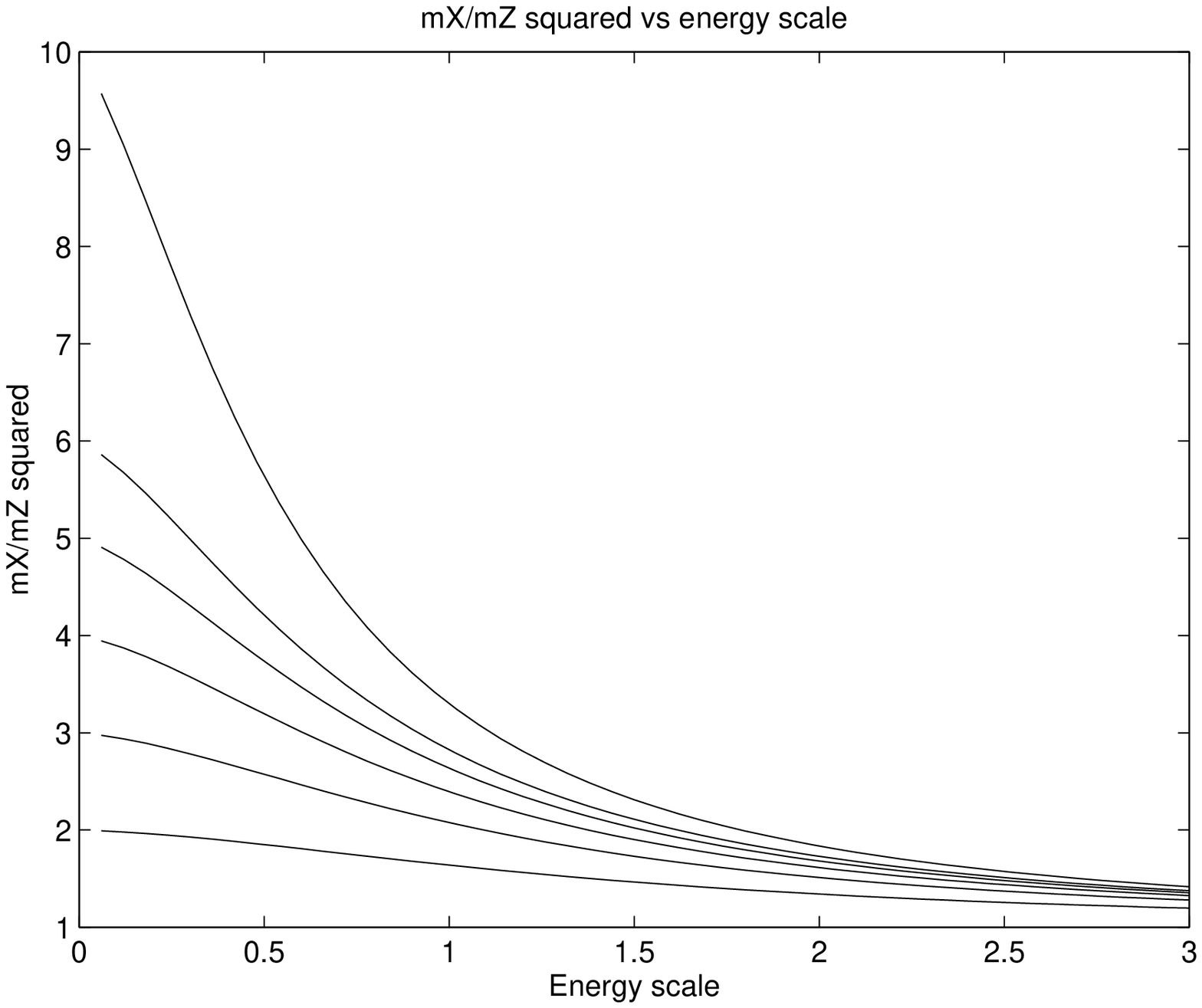}}
\in
{\it \noindent Fig.5:
Evolution of the ratio $(m_X / m_Z )^2$ from $M_P$ to
$M_U$, for various input values at $M_P$. The $x$-axis is
$log_{10}(M_P / {\mu})$}.
\medskip
\out

In all the above results it is assumed that there is no  dependence on
the  flavour indices $\alpha, \beta \cdots$. We have explored
various specific forms for  flavour dependence, and found that
whether IR stability is maintained depends on the flavour
structure. Before an exhaustive analysis of the possibilities
it may be appropriate to construct a more realistic theory.

In this model there are no gauge invariant $b_{ij}$ terms. For models
with such terms, it  is our  experience that it is typically difficult
to arrange for both  IR stability of Eq.~\Ek{c}\ and for rapid approach
to the fixed point.  From this point of view, the relevant boundary
conditions for low  energy phenomenology may be $-A = \sqrt{3}m = M$,
with $B$ a free  parameter, rather than $B = 2 M /\sqrt{3}$ as in
Refs.~\jjra , \jbalt .  Precisely these boundary conditions were in fact
explored in  Ref.~\ref\blm{R.~Barbieri, J.~Louis and M.~Moretti, \plb
312 (1993) 451; erratum--{\it ibid\/} 316 (1993) 632\semi  J.L.~Lopez,
D.V.~Nanopoulos and A.~Zichichi, \plb 319 (1993) 451}. We would  argue
that they are relevant without the need of the special assumptions
necessary to derive them from string theory.

We conclude by a reiteration of our basic philosophy, which transcends
the details of the toy models we have presented. If universal scalar
masses  and cubic couplings at $M_U$ are to be an infra--red
phenomenon, they will  necessarily be of the specific form shown in
Eq.~\Ek{a}\ and \Ek{b},  and this can only be achieved in the class of
theories which  can satisfy $P = {1\over 3}Q$. This results in a
substantial  sharpening of the predictions for the superparticle mass
spectrum at low  energies\blm ; even more so\jjra\  if we suppose that
Eq.~\Ek{c} is also approached.

\bigskip\centerline{{\bf Acknowledgements}}\nobreak

IJ was supported by PPARC via an Advanced Fellowship, and PF by a
scholarship from JNICT.

\listrefs
\bye